\documentstyle[prd,aps,psfig]{revtex}
\def\simlt{\stackrel{<}{{}_\sim}}
\def\simgt{\stackrel{>}{{}_\sim}}
\newcommand\be{\begin{equation}}
\newcommand\ee{\end{equation}}
\newcommand\bea{\begin{eqnarray}}
\newcommand\eea{\end{eqnarray}}
\newcommand\ba{\begin{array}}
\newcommand\ea{\end{array}}
\begin{document}
\draft
\input epsf
\def\la{\mathrel{\mathpalette\fun <}}
\def\ga{\mathrel{\mathpalette\fun >}}
\def\fun#1#2{\lower3.6st\vbox{\baselineskip0st\lineskip.9st
        \ialign{$\mathsurround=0pt#1\hfill##\hfil$\crcr#2\crcr\sim\crcr}}}

\twocolumn[\hsize\textwidth\columnwidth\hsize\csname
@twocolumnfalse\endcsname

\title{Massive Neutrinos and the Higgs Mass Window}
\author{J.A. Casas$^{(1, 2)}$, V. Di Clemente$^{(2)}$, A. Ibarra$^{(2)}$ 
and M. Quir\'os$^{(2)}$}
\address{\phantom{ll}}
\address{$^{(1)}${\it CERN TH-Division, CH-1211 Geneva 23, Switzerland}}
\address{$^{(2)}${\it IEM, CSIC Serrano 123, 28006 Madrid, Spain}}
\date{\today}
\maketitle
\begin{abstract}
If neutrino masses are produced by a see-saw mechanism
the Standard Model prediction for the Higgs mass window (defined by  
upper (perturbativity) and lower (stability) bounds) can be substantially 
affected. Actually the Higgs mass window can close completely, 
which settles an upper 
bound on the Majorana mass for the right-handed neutrinos, $M$, ranging from 
$10^{13}$ GeV for three generations of quasi-degenerate 
massive neutrinos with $m_\nu\simeq 2 $ eV, to $5\times 10^{14}$ GeV for 
just one relevant generation with $m_\nu\simeq 0.1 $ eV. A slightly weaker 
upper bound on $M$, coming from the requirement that the neutrino Yukawa 
couplings do not develop a Landau pole, is also discussed. 
\end{abstract}
\pacs{PACS: 14.80.Bn, 14.60.Pq, 14.60.St~
CERN-TH/99-93~ IFT-UAM/CSIC-99-14~
IEM-FT-190/99\\
{\sf hep-ph/9904295}}
\vskip2pc]

Observations of the flux of atmospheric neutrinos by 
SuperKamiokande~\cite{SK}
provide strong evidence for neutrino oscillations, which in turn imply
that (at least two species of) neutrinos must be massive.  Additional
support to this hypothesis is given by the need of neutrino
oscillations to explain the solar neutrino flux deficit and
the possible essential role of the neutrinos in the large scale
structure of the universe~\cite{kayser}.  
Much work has been devoted in the
last months in order to guess and to explain the structure and the origin
of the neutrino mass matrices capable to account for the different
observations~\cite{work}.

In this letter, we would like to point out the consequences
that massive neutrinos have for the Standard Model (SM), in particular
for its still unprobed sector, namely the Higgs sector. As
it is known, the Higgs mass in the SM is bounded from above from the
requirement of validity of perturbation theory,
and from below by the
requirement of stability of the Higgs potential~\cite{stab0,stab1}. 
More precisely,
if we demand both requirements until the Planck scale ($M_{P\ell}$), then the
allowed window for the Higgs mass is
\be
137\ {\rm GeV} \simlt M_H \simlt 175\ {\rm GeV}\ .
\label{SMwindow}
\ee
We will show that if neutrino masses are produced
by a see-saw mechanism, the previous prediction for
the Higgs mass can be substantially modified.

The simplest extension of the SM Lagrangian, capable to account for
neutrino 
masses, reads
\be
{\cal L}_\nu = -\bar \nu_R {\cal M}_D \nu_L 
- \frac{1}{2} \bar \nu_R {\cal M}_M \bar \nu_R^T + \ {\rm h.c.}\ , 
\label{Lnu}
\ee
where ${\cal M}_D$ is the Dirac mass matrix (${\cal M}_D = \frac{1}{\sqrt2}
\langle \phi \rangle {\bf Y_\nu}$, where $\phi$ is the neutral
component of the Higgs 
field and ${\bf Y_\nu}$ is the matrix of the neutrino Yukawa couplings) 
and ${\cal M}_M$ is the Majorana mass matrix for the right-handed neutrinos.

In order to discuss the impact of the previous extension of the SM on
the Higgs sector, it is convenient to start with the case where there
is a hierarchy of left-handed neutrino masses,  $m_{\nu_1}^2\ll
m_{\nu_2}^2\ll m_{\nu_3}^2$, presumably inherited by a similar
hierarchy in the Dirac-Yukawa couplings. In this case, there is only
one relevant generation of neutrinos for our purposes 
(the most massive one), so 
${\cal M}_D$ and ${\cal M}_M$ become the single parameters
$m_D$ and $M$. As we will see the
extension of the results to the general case is straightforward.  Now,
the two associated neutrino eigenvalues arising from (\ref{Lnu}) are
\bea
m_{\nu_{1,2}} = \frac{1}{2}\left(M \mp \sqrt{M^2+4m_D^2}\right)
\label{mnus}
\eea
For $M\gg\langle\phi\rangle$, we get $m_{\nu_1}\simeq m_D^2/M$,
$m_{\nu_2}\simeq M$; $\nu_1, \nu_2$ correspond essentially to the
left- and right-handed neutrinos respectively. For energy scales below $M$ 
we can integrate out the $\nu_2$ neutrino, so ${\cal L}_\nu$ effectively 
becomes
\bea
{\cal L}_\nu^{\rm eff} 
&=& -\frac{1}{2} \nu_L^T m_{\nu_1} \nu_L + {\rm h.c.} 
\equiv -\frac{1}{4} \kappa \nu_L^T \langle\phi\rangle^2 \nu_L+ {\rm h.c.}
\label{Lnueff}
\eea
where $v=\langle\phi\rangle$ at the physical vacuum (i.e. $v=246$ GeV) and 
for convenience we have introduced the effective coupling $\kappa$
($\kappa v^2=2\,{m_D^2}/{M}$ at the $M$-scale).
For scales below $M$, $\kappa$ is a running parameter, whose beta function 
is essentially given by~\cite{babu}
\bea
\beta_\kappa=\frac{1}{16\pi^2}\left(-3g_2^2+2\lambda+6Y_t^2\right)\kappa\
,
\label{betakapa}
\eea
where $g_2,\lambda,Y_t$ are the $SU(2)$ gauge coupling, the Higgs
quartic coupling and the top Yukawa coupling respectively. Thus, for a
given physical mass of $\nu_1$ (to be identified with the low-energy value
of $\kappa v^2/2$) and for a given value of the Majorana mass, $M$, the
Dirac mass $m_D$ (and thus $Y_\nu$) is unambiguously fixed. 

The one-loop effective potential $V(\phi)$ has the form 
\bea
V=V_{SM}+\Delta V_\nu
% V=-\frac{1}{2}m^2 + \frac{1}{8}\lambda \phi^4 + \Omega + \Delta
% V^{1-loop}\ ,
\label{V}
\eea
where $V_{SM}$ is the usual one-loop SM potential, consisting of the
tree level part
$V_{tree}=-\frac{1}{2}m^2\phi^2+\frac{1}{8}\lambda\phi^2+\Omega$, plus
the ordinary radiative corrections dominated by the top Yukawa coupling, $Y_t$,
and 
\bea
\Delta V_\nu&=&-\frac{1}{32\pi^2}\left[m_{\nu_1}^4\log\frac{m_{\nu_1}^2}
{\mu^2}+\theta_\nu\, m_{\nu_2}^4\log\frac{m_{\nu_2}^2}{\mu^2}
\right]\ ,
\label{DeltaV}
\eea
where $m_{\nu_{1,2}}$ are given by (\ref{mnus}) (note that they are
functions  of $\phi$), $\mu$ is the renormalization
scale\footnote{More precisely, $\mu^2=e^{3/2}\bar\mu^2$, where $\bar
\mu$ is the usual $\overline{\rm MS}$ renormalization scale.} and 
$\theta_\nu\, \equiv\theta(\mu-M)$~\cite{japs,scales}
is a step $\theta$-function accounting for the threshold at the $M$-scale.

Above $M$, $Y_\nu$ runs with the scale, while the various SM
parameters get contributions from $Y_\nu$ to their
beta-coefficients. The relevant ones are~\cite{RGE}
%%
%
%also contributions from the neutrino
%Yukawa coupling. The relevant $\nu$-contributions to the beta-coefficients 
%are~\cite{RGE}
%
\bea
\beta_{Y_\nu}&=&\frac{1}{16\pi^2}\ \theta_\nu\, \left[3 Y_t^2-
\left(\frac{3}{4}g_1^2
+\frac{9}{4}g_2^2\right)+\frac{5}{2}Y_\nu^2
\right] Y_\nu \nonumber\\
\beta_\lambda^{(\nu)}&=&\frac{1}{16\pi^2}\left[4\theta_\nu 
(\lambda Y_\nu^2-Y_\nu^4)\right],
\; \gamma_{\phi}^{(\nu)}=
\frac{1}{16\pi^2}\left[\theta_\nu Y_\nu^2\right]\hspace{-0.8cm}
 \nonumber\\
\beta_{\Omega}^{(\nu)}&=&-\frac{1}{16\pi^2}\left[\theta_\nu M^4 \right],
\;\;\;\beta_{m^2}^{(\nu)}
=\frac{1}{16\pi^2}\left[4\theta_\nu M^2 Y_\nu^2\right]
\nonumber\\
\beta_{Y_t}^{(\nu)}&=&\frac{1}{16\pi^2}\left[\theta_\nu Y_\nu^2Y_t\right],
\label{betas}
\eea
where $\beta_i^{(\nu)}$ is the contribution of $Y_\nu$ to the $\beta_i$
coefficient.
The matching of the complete and the effective theory at $M$ requires
to introduce threshold corrections. In particular, the matching of $V$
requires to introduce a threshold contribution below $M$:
$\Delta_{th}V=-\frac{2}{64\pi^2}\left[m_{\nu_2}^4\log\frac{m_{\nu_2}^2}{M^2}
\right]$, whose expansion gives the threshold corrections to the  $m^2$
and $\lambda$ parameters, namely $\Delta_{th}m^2=
\frac{1}{16\pi^2}Y_\nu^2M^2$ \footnote{Note that the threshold
correction for $m^2$ is very sizeable, exhibiting an explicit
implementation of the SM gauge hierarchy problem. Working within the
SM framework, it is necessary to fine-tune the $m^2$ at high energy
in order to reproduce the usual low-energy physics.  However, this
conceptual shortcoming does not play any role in our analysis and
the corresponding results, since the $m^2\phi^2$ contribution to the
effective potential is always negligible compared to the
$\lambda\phi^4$ one.},  $\Delta_{th}\lambda= -\frac{5}{16\pi^2}Y_\nu^4$.

Now we are prepared to compute the modification of the perturbativity
and stability bounds on the SM Higgs mass. The perturbativity bound
arises from the requirement that $\lambda$ does not enter the
non-perturbative regime below a certain scale $\Lambda$. 
If the SM is to be valid until
$M_{P\ell}$ or some high scale $M_X$, $\Lambda$ should be
identified with those scales. Since for $\mu<M$ the running of $\lambda$
is as in the SM, the perturbativity bound for $\Lambda<M$
remains the same. However, for $\mu>M$ the contribution of the
neutrinos to the $\lambda$ running (which for large $\lambda$ is
dominated by the positive $\lambda Y_\nu^2$ term in eq.~(\ref{betas})) 
speeds up the increasing of $\lambda$, thus leading to a more stringent 
upper bound on $M_H$.

On the other hand, the stability bound on the Higgs mass arises from
demanding that the potential does not develop an instability
for large values of $\phi$. The
instability arises from the fact that $\lambda$ can be
driven to negative values at large enough scales.  Notice that to
evaluate the potential for large values of $\phi$ one has to plug a
renormalization scale $\mu\sim \phi$ in order to avoid large
logarithms. Hence, if the $\lambda\phi^4$ term, which is dominant for
large $\phi$, becomes negative, so does the potential. If we demand
this not to happen below a scale $\Lambda$, i.e. we require
$\lambda(\Lambda)\geq 0$, this translates into a lower bound 
on $M_H$. Again, for $\Lambda<M$
the values of $\lambda$ are as in the SM, and thus the corresponding
lower bound on $M_H$. However, for $\Lambda>M$ the values of $\lambda$
get modified by the neutrino contribution to $\beta_\lambda$. For
small values of $\lambda$ this contribution is dominated by the
negative $Y_\nu^4$ term in (\ref{betas}). Hence, $\lambda$ is driven
more rapidly to negative values and the lower bound becomes more
stringent too.

For a more careful analysis of the stability bound, one has to
consider not just the tree-level potential (which is in fact dominated
by the $\lambda\phi^4$ term), but also the radiative corrections.
A practical way to include them~\cite{stab1} is to
take $\mu=\alpha \phi$ ($\alpha\simeq 1$ is always a correct choice) and
then to extract the $\sim \phi^4$ contributions from $\Delta
V^{1-loop}$, neglecting (for $\phi\gg M$) the $M/\phi$ factors.  These
$\phi^4$ contributions can be incorporated into an effective quartic
parameter:
\bea
\lambda_{e\hspace{-0.5mm}f\hspace{-0.5mm}f}
(\mu<M) &=& \lambda -\frac{1}{16\pi^2}\left[ 6Y_t^4\log\frac{Y_t^2}{2}
\right]
 \nonumber\\
\lambda_{e\hspace{-0.5mm}f\hspace{-0.5mm}f}
(\mu>M) &=& \lambda -\frac{1}{16\pi^2}\left[ 6Y_t^4\log\frac{Y_t^2}{2}
+2Y_\nu^4\log\frac{Y_\nu^2}{2}
\right]
\label{leff}
\eea
Notice that for $\mu<M$ the $\nu_1$ contribution (the only neutrino
which is present) is negligible, as $m_{\nu_1}\ll m_t$. However, for
$\mu>M$ both neutrinos contribute in the same (non-negligible) amount.
The stability bound reads, in this more accurate form, as
$\lambda_{e\hspace{-0.5mm}f\hspace{-0.5mm}f}(\Lambda)\geq 0$.

To summarize, the presence of the neutrinos strengthens both the upper
and the lower bound on the SM Higgs mass, thus narrowing the Higgs
window.  The quantitative effect depends just on the value of the
``left-handed'' neutrino mass, $m_{\nu_1}$, and on the value of the
Majorana mass, $M$ (the Yukawa coupling, $Y_\nu$, can be extracted from
those two in the way explained above).

Before presenting numerical results, let us extend the analysis to the
case where all the neutrinos have similar masses, and therefore all of
them are equally relevant. Then, $\kappa$, ${\cal M}_M$, ${\cal M}_D$ and 
${\bf Y_\nu}$ are flavor
matrices. The corresponding effective mass matrix for the left-handed
neutrinos (i.e. the analogue to $m_{\nu_1}\simeq m_D^2/M$ in the
single neutrino case) is given by
\bea
{\cal M}_\nu={\cal M}_D^T {\cal M}_M^{-1} {\cal M}_D
\label{Mnu}
\eea
The beta functions given in eqs.~(\ref{betas}) get slightly modified, depending
now on the whole matrix ${\bf Y_\nu}$ (the extension is quite trivial
and can be found in~\cite{RGE}).  Consequently, the precise results
will depend in principle on the textures of the ${\cal M}_D$ 
(or ${\bf Y_\nu}$) and ${\cal M}_M$
matrices. However, as we will see, in practice the form of the
textures is not important in our analysis. 

One can always choose a
basis of $\nu_{L_i},\nu_{R_i}$ in which $m_D$, and thus ${\bf Y_\nu}$, 
is diagonal (in
this basis the matrix of Yukawa couplings of the charged leptons  is
in general non-diagonal, but these couplings are negligible in our
analysis). Then, the relevant formulas, eqs.~(\ref{DeltaV})-(\ref{leff}) 
become trivially
extended to the 3-generation case~\cite{RGE}.
%(each neutrino would have exactly 
%a separate contribution of the form expressed in those equations).
%
If the physical neutrinos are
quasi-degenerated (which, according to the observations must occur for
$m_{\nu_i}\simgt 0.1 eV$), it is hard to believe that this fact
can naturally
come from a diagonal ${\cal M}_D$ with non-degenerated entries, compensated 
by a non-trivial structure of the ${\cal M}_M$ matrix, see eq.~(\ref{Mnu}).
One can arrange things in this way, but it is extremely artificial.
(In this case, the largest Yukawa coupling would 
dominate all the equations, and we simply would be back to the case 
of a single relevant neutrino). Therefore, one expects that both
${\cal M}_D$ and ${\cal M}_M$ have essentially degenerated eigenvalues,
$m_D$ and $M$, so that
at the end of the day all the neutrinos have masses 
$m_{\nu_i}\simeq m_D^2/M$. 

On the other hand, it has been claimed in the literature~\cite{GG}
that if neutrino masses are to play an essential cosmological role, 
beside explaining the atmospheric and solar neutrino 
fluxes, the texture of the 
effective left-handed neutrino mass must be of the bimaximal mixing type
\bea
{\cal M}_\nu=m_\nu \left(\begin{array}{ccc} 0&1/\sqrt{2}&1/\sqrt{2}\\
1/\sqrt{2}&1/2&-1/2\\
1/\sqrt{2}&-1/2&1/2
\end{array}\right)
\label{MGG}
\eea 
where $m_\nu$ is the common neutrino mass. This texture produces
three degenerate left-handed neutrinos with maximal mixing angles
$\theta_{12}$ and $\theta_{23}$. This is the texture we have used for
our numerical analysis, although, as explained above, other 
textures will produce similar results.
Notice that for our analysis we need not just the form of 
${\cal M}_\nu$, but also the form of the matrix of the Yukawa couplings,
${\bf Y_\nu}$, and thus the form of ${\cal M}_M$ (see eq.~(\ref{Mnu})).
Our choice has been to take ${\bf Y_\nu}$ proportional to the 
identity, i.e ${\bf Y_\nu} = Y_\nu\ {\rm diag}(1,1,1)$.
As discussed above, this is a perfectly reasonable choice in this case.
Then the form of the right-handed Majorana matrix is forced to be
\bea
{\cal M}_M^{-1}=\frac{1}{Mm_\nu}{\cal M}_\nu \ .
\label{MMGG}
\eea
It is straightforward to check that
with these ansatzs, the mass-squared eigenvalues of the three
``left-handed'' and the three ``right-handed'' 
neutrinos, say $(m_{\nu_{1,2}}^{(i)})^2$, with $i=1,2,3$,
are as in eq.~(\ref{mnus}). 
Thus, we see that in this particularly well-motivated case the 
texture plays no significant role indeed.
%%
%Alternatively, on can always choose a basis of $\nu_{R_i}$ in which
%${\cal M}_M\equiv M\,{\rm diag}(1,1,1)$ 
%is diagonal, in which case there is no
%accompanying rotation in the charged lepton sector. The relevant term in the
%$\beta^{(\nu)}_\lambda$ coefficient is readily seen to be, 
%${\rm Tr}({\bf Y_\nu}^T {\bf Y_\nu})^2=3\,\left(4M^2m_\nu^2/v^4\right)
%\equiv 3\, Y_\nu^4$, where now $Y_\nu$ is the Yukawa coupling corresponding
%to a single generation of neutrinos with mass $m_\nu$. Therefore the 
%modification of the stability bound is similar to that corresponding to three
%degenerate neutrinos with mass $m_\nu$, and again the texture plays no
%relevant role.
%

Let us now present the results. Fig.~1 shows the evolution of the 
Higgs window as a function of the scale $\Lambda$ until which the 
theory is valid. In order to illustrate the effect of the neutrinos 
we have chosen a typical case, namely $m_\nu=0.1$ eV (for one or 
the three generations of neutrinos) and $M\simeq 3\times 10^{14}$ GeV.
The pure SM window is also shown to facilitate the comparison.
We note that the strengthening of the stability (lower) bound is 
the main responsible of the substantial narrowing of the allowed window.

%%%%%%%%%%%%%%%%%%%%%%%%figure%%%%%%%%%%%%%%%%%%%%%%%%
\begin{figure}[hbt]
%%\psdraft   
\centerline{
\psfig{figure=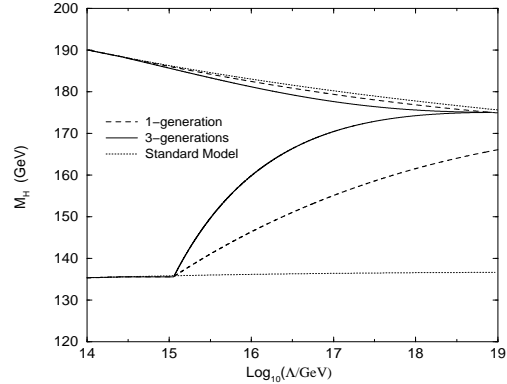,height=5.cm,width=7.cm,bbllx=2.cm,bblly=1.cm,bburx=21.cm,bbury=15.cm}}
\caption{\footnotesize The Higgs window as a function of the scale
$\Lambda$
for $m_\nu = 0.1$ eV and $M = 2.8\times 10^{14}$ GeV.}
\end{figure}
%%%%%%%%%%%%%%%%%%%%%%%%figure%%%%%%%%%%%%%%%%%%%%%%%%

Fig.~2 shows the variation of the Higgs window with the Majorana mass,
$M$, for $\Lambda=M_{P\ell}$ (continuous lines) and $\Lambda=10^{16}$ GeV 
(dashed lines) for
just one relevant generation of massive neutrinos with $m_\nu = 0.1$
eV (the other generations may be massive but hierarchically smaller). 
%
%%%%%%%%%%%%%%%%%%%%%%%%figure%%%%%%%%%%%%%%%%%%%%%%%%
\begin{figure}[hbt]
%%\psdraft   
\centerline{
\psfig{figure=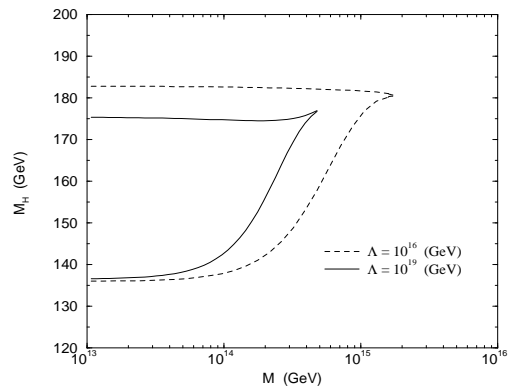,height=5.cm,width=7.cm,bbllx=2.cm,bblly=1.cm,bburx=21.cm,bbury=15.cm}}
\caption{\footnotesize The Higgs window as a function of the Majorana mass,
$M$ for $\Lambda=M_{P\ell}$ (continuous lines) and $\Lambda=10^{16}$ GeV 
(dashed lines) for the case of just one massive neutrino with 
$m_\nu = 0.1$ eV.}
\end{figure}
%%%%%%%%%%%%%%%%%%%%%%%%figure%%%%%%%%%%%%%%%%%%%%%%%%
%
\noindent
This case corresponds to the most conservative scenario 
concerning the effect of neutrinos on the Higgs window.
The values of the bounds for ``low'' Majorana masses
($M\simlt 10^{13}$ GeV) coincide with those of the pure SM without
neutrinos. Fig.~3 is analogous to Fig.~2, but for three generations of 
neutrinos, with
degenerate masses $m_\nu = 2$ eV. This case corresponds to the
scenario where the effect of the neutrinos in the Higgs sector is
maximized. 
%%%%%%%%%%%%%%%%%%%%%%%%figure%%%%%%%%%%%%%%%%%%%%%%%%
\begin{figure}[hbt]
%%\psdraft   
\centerline{
\psfig{figure=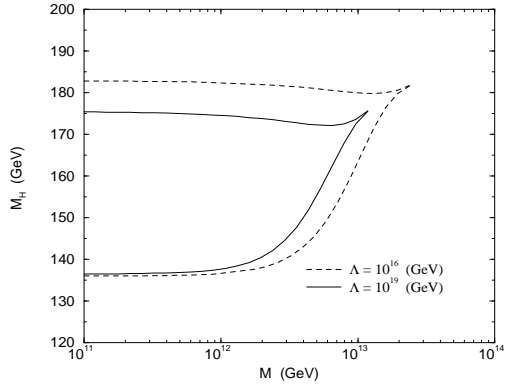,height=5.cm,width=7.cm,bbllx=2.cm,bblly=1.cm,bburx=21.cm,bbury=15.cm}}
\caption{\footnotesize The same as Fig.1, but with three generations of
massive neutrinos with  $m_\nu = 2$ eV.}
\end{figure}
%%%%%%%%%%%%%%%%%%%%%%%%figure%%%%%%%%%%%%%%%%%%%%%%%%
\noindent
It is apparent from Figs.~2 and~3 that above a certain value of the
Majorana mass the Higgs window closes up, disappearing. In principle,
this could seem paradoxical, since one expects that for some
fine-tuned initial value of $\lambda$ (and thus of $M_H$) the
evolution of $\lambda(\mu)$ for large  $\mu$ will be just between the
two bounds, say  $0\leq\lambda(\mu)\leq 4\pi$.
So one would expect,
for large values of $M$, an extremely narrow window, but a window after
all.  The fact that actually forbids that window is that for large
values of $M$ the neutrino Yukawa couplings themselves develop a
Landau pole below $\Lambda$. 
Alternatively, we can set
$Y_\nu$ at the Landau pole at $M_{P\ell}$
(i.e. $Y_\nu(M_{P\ell})\gg 1$) and evaluate the corresponding
low energy value of $m_\nu$, through the renormalization group equations
(RGE) of $Y_\nu$ and $\kappa$,
for a certain value of the Majorana mass $M$. 
%%%%%%%%%%%%%%%%%%%%%%%%figure%%%%%%%%%%%%%%%%%%%%%%%%
\begin{figure}[hbt]
%%\psdraft   
\centerline{
\psfig{figure=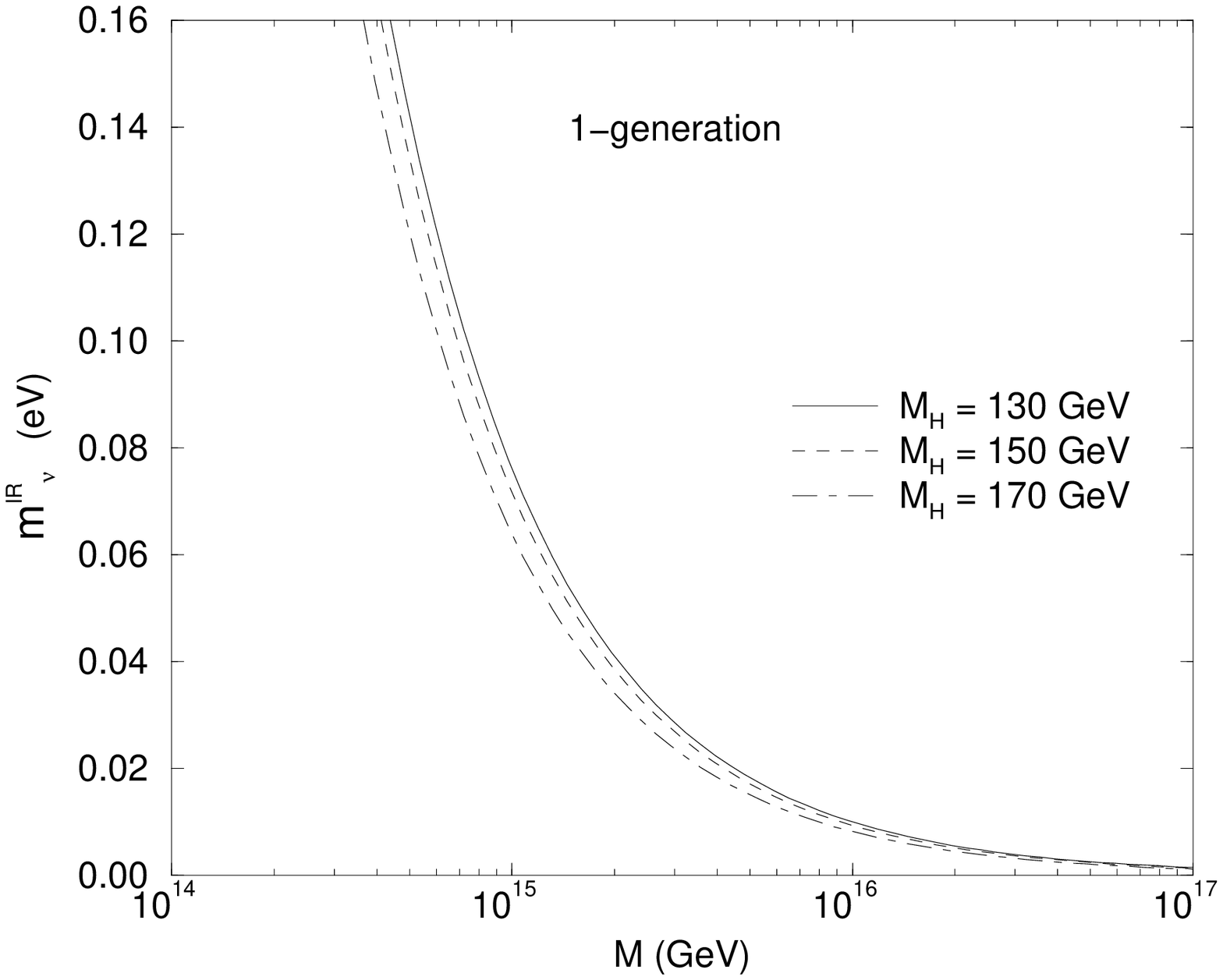,height=4.5cm,width=4.5cm,angle=0,bbllx=0.cm,bblly=1.5cm,bburx=19.cm,bbury=15.5cm}
\psfig{figure=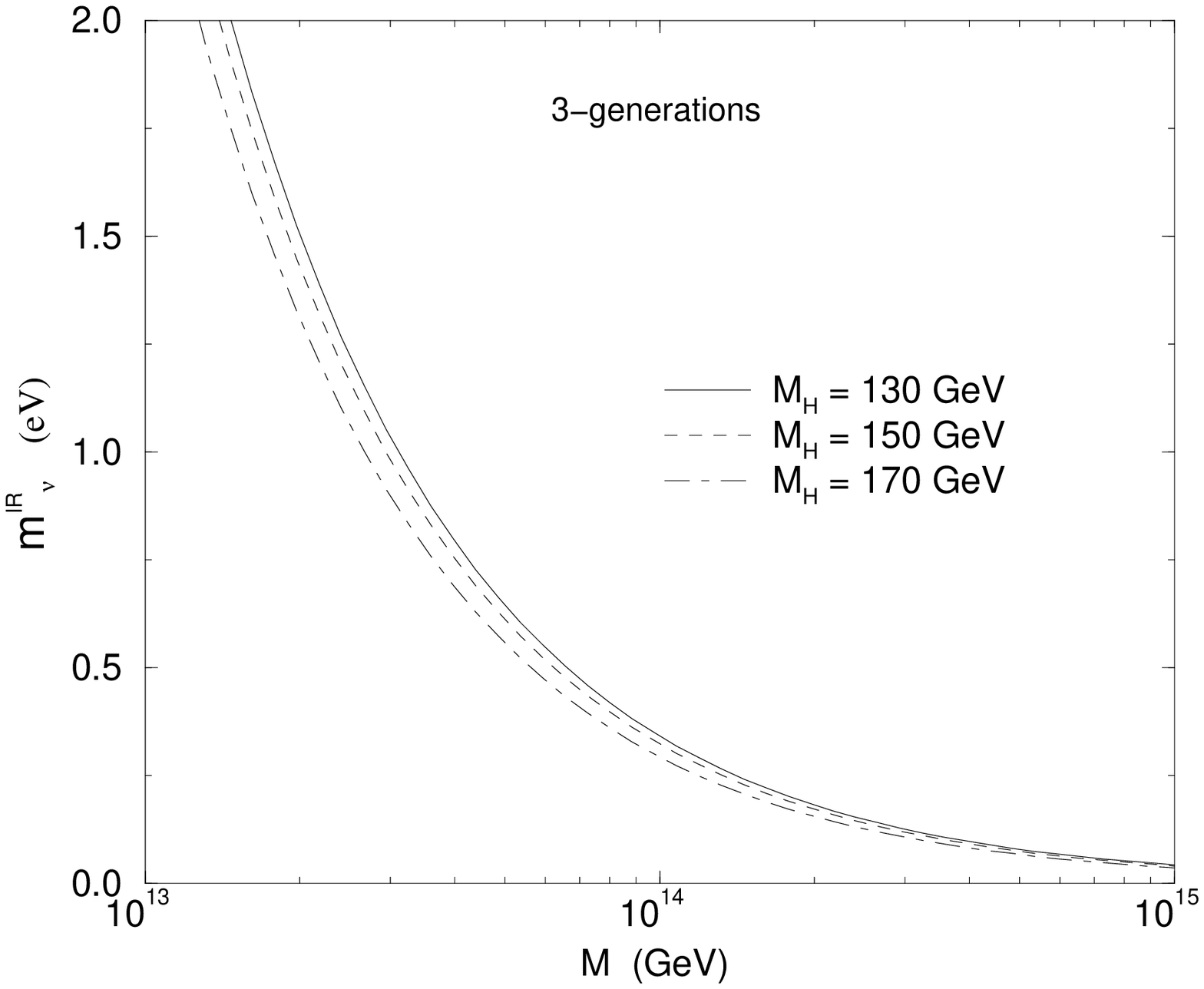,height=4.5cm,width=4.5cm,angle=0,bbllx=2.cm,bblly=1.5cm,bburx=21.cm,bbury=15.5cm}}
\caption{\footnotesize Upper bound on the neutrino mass, $m_\nu^{IR}$,
vs. the Majorana mass $M$ for one and three generations of massive neutrinos
(left and right panels respectively).}
\end{figure}
%%%%%%%%%%%%%%%%%%%%%%%%figure%%%%%%%%%%%%%%%%%%%%%%%%
\noindent
This ``infrared fixed point''
value, say $m_\nu^{IR}$, represents an upper bound for the neutrino mass.
The dependence of $m_\nu^{IR}$ on $M$ is illustrated in Fig.~4 for one 
and three generations of massive neutrinos. Due to the dependence of the
$\kappa$ RGE on $\lambda$, the value of $m_\nu^{IR}$ presents a slight
dependence on the value of $M_H$, as is shown in the figure.

In conclusion, we have shown that if neutrino masses are produced
by a see-saw mechanism, the SM prediction for
the Higgs mass window (defined by upper (perturbativity) and  
lower (stability) bounds)
is substantially affected in an amount that depends on
the value of the Majorana mass for the right-handed neutrinos, $M$. Actually,
for values of $M$ above a certain value, the Higgs window closes,
setting an upper bound on $M$. This varies from $10^{13}$ GeV
for three generations of massive neutrinos with $m_\nu\simeq 2 $ eV
to $5\times 10^{14}$ GeV for just one relevant generation with
$m_\nu\simeq 0.1 $ eV.
We have also discussed a
second (slightly weaker) upper bound on $M$, coming from the
requirement that the neutrino Yukawa couplings do not develop 
a Landau pole. The whole analysis and results are practically
independent of the details of the model (i.e. the particular structure
of the neutrino mass matrices).

\end{document}